\documentstyle[12pt]{article}
\makeatletter

\ifcase \@ptsize
\or 
\or 
\fi
\advance\textheight by \topskip

\ifcase \@ptsize
\or 
\or 
\fi

\renewcommand{\theequation}{\@arabic\c@section~.~\@arabic\c@equation}
\def\section{\c@equation=0%
\@startsection {section}{1}{\z@}{-3.5ex plus -1ex minus
 -.2ex}{2.3ex plus .2ex}{\large\bf\centering}}
\def\subsection{\@startsection{subsection}{2}{\z@}{-3.25ex plus -1ex minus
 -.2ex}{1.5ex plus .2ex}{\sc}}

\def\@cite#1#2{\nolinebreak$^{[\scriptstyle #1\if@tempswa , #2\fi]}$}
\def\@citex[#1]#2{\if@filesw\immediate\write\@auxout{\string\citation{#2}}\fi
  \def\@citea{}\@cite{\@for\@citeb:=#2\do
    {\@citea\def\@citea{,\penalty\@m}\@ifundefined
       {b@\@citeb}{{\bf ?}\@warning
       {Citation `\@citeb' on page \thepage \space undefined}}%
{\csname b@\@citeb\endcsname}}}{#1}}
\@definecounter{footnote}

\gdef\@publabel{\hfil}
\gdef\@pubdate{\null}
\gdef\@pubnumber{\null}
\gdef\@author{\null}
\gdef\@title{\null}
\gdef\@abstract{\null}
\long\def\pubdate#1{\gdef\@pubdate{#1}}
\long\def\pubnumber#1{\gdef\@pubnumber{#1}}
\long\def\publabel#1{\gdef\@publabel{#1}}
\long\def\author#1{\gdef\@author{#1}}
\long\def\title#1{\gdef\@title{#1}}
\long\def\abstract#1{\gdef\@abstract{#1}}
\def\titlerelax{
\let\maketitle\relax
\let\settitleparameters\relax
\let\consolidatetitle\relax
\let\inittitlepage\relax
\let\finishtitlepage\relax
\let\titlepagecontents\relax
\let\multithanks\relax
\let\titlebaselines\relax
\let\@makepub\relax
\let\@maketitle\relax
\let\@makeauthor\relax
\let\@makeabstract\relax
\let\@maketitlenote\relax
\let\thanks\relax
\let\titlerelax\relax}
\def\titleclean
{\gdef\@titlenote{}
\gdef\@abstract{}
\gdef\@author{}
\gdef\@title{}
\gdef\@pubdate{}\gdef\@pubnumber{}\gdef\@publabel{}
\gdef\@dpublabel{}
}

\def\@makepub{\vbox to \z@{\hbox to \textwidth{\hfill
\@publabel \hfill
\llap{\parbox[t]{0.25\textwidth}{\raggedleft\@pubnumber}}}%
\vss}}
\def\@maketitle{\vskip 80pt \begin{center}
 {\LARGE \@title \par}
 \end{center}}
\def\@makeauthor{{\def\and{\smallskip {\normalsize \rm and\smallskip}}
\long\def\address##1{{\def\and{\\and\\}\medskip
				{\small \it \\##1\\}
}}
{\centering
 \vskip 2.5em
 \large \lineskip .75em
 \@author}
 \par}}
\def\@makedate{\vskip 1.5em
 {\raggedright \small \noindent\@pubdate \par}}
\def\@makeabstract{\vskip 1.5em
{\small
\begin{center}
{\bf ABSTRACT\vspace{-.5em}\vspace{0pt}}
\end{center}
\quotation \@abstract \endquotation}}
\def\maketitle{
\let\footnotesize\small \setcounter{page}{1}
\@makepub
\@maketitle
\@makeauthor
\@makeabstract
\@thanks
\@makedate
\setcounter{footnote}{0}
}

\begin{document}
\newcommand{\hsp}{\hspace{0.08in}}
\newcommand{\eqn}{\begin{equation}}
\newcommand{\e}{\end{equation}}
\bibliographystyle{npb}

\pubnumber{ UMTG - 243 \\\today}

\title{ Factorization Method and the Supersymmetric Monopole Harmonics}
     
 \author{G. Andrei Mezincescu\footnote{On leave of absence from INFM, C.P. MG-7,
         R-76900 M\u agurele, Ilfov, Rom\^ania, and Centrul de Studii
         Avansate de Fizic\u a al Academiei Rom\^ane, Bucure\c sti,
         Rom\^ania.amezin@ctsps.cau.edu}\linebreak   
  \address {C.T.S.P.S.,\linebreak 
             Clark-Atlanta University, \linebreak
             Atlanta, GA 30314 \linebreak
             }   
             Luca Mezincescu\footnote{mezincescu@phyvax.ir.miami.edu}\linebreak  
   \address{Department of Physics, \linebreak
              University of Miami, \linebreak
              Coral Gables, Fl.33124 \linebreak
               }}

\abstract{
 We use the general $ N = 1$ supersymmetric formulation of one 
 dimensional sigma models on non trivial manifolds and its 
 subsequent quantization to formulate the classical and quantum
 dynamics of the $ N= 2 $ supersymmetric charged particle moving 
 on a sphere in the field of a monopole. The factorization method
 is accommodated with the general covariance and it is used to integrate
 the corresponding system.
}

\maketitle
\baselineskip 18pt
\parskip 18pt
\parindent 10pt
\vfill\pagebreak
\section{Introduction }

The factorization method was first used by Schr{\H{o}}dinger \cite{S} to 
diagonalize the harmonic oscillator. It may work when the
Hamiltonian of the system can be cast as a product of two operators: 
\eqn\label{Hfactor}
 H = AB\;,
\e 
a c-number can also be added to the above operator.

Most systems treated by this method are one dimensional systems 
 \cite{inf}  \cite {shabat}, and it  has 
been also considered for other situations\cite {sabatier}. 

Recently, Ferapontov and Veselov \cite{fera} used  factorization to look 
for integrable Schr{\H{o}}dinger operators with magnetic fields on 
two dimensional surfaces. In the following we will use only their solution
for the monopole harmonics of Wu and Yang  \cite{ttwu}.

We will show that factorization can be readyly used for integrating the 
$ N= 2 $ supersymmetric nonrelativistic quantum mechanics \cite{ed} of a
particle 
with spin 
moving on a sphere in the field of a monopole placed at its center.
This can be inferred from the fact that the supersymmetric charge of
the classical $\sigma$ - model quantizes as the Dirac 
operator\cite{louis} \cite{dan} on the 
manifold. Because the sphere is a two dimensional manifold, in 
a convenient basis the Dirac operators do not mix the components of the 
spinors and therefore the  Hamiltonian of the system will be:
$$ 
H = {1\over 2} ( Q\bar Q + {\bar Q} Q ) =
\left( \begin{array}{cc}
   AB   &   0\\
   0   &   BA
   \end {array} 
   \right )\;, 
$$
and because of this, the factorization method will be the natural method to 
integrate $ H $.

As the factorization method is very   simple we will replay some 
of its features here, in order to make the following computations easier
to follow. 
 Therefore along with  (\ref {Hfactor}), one considers the reverse order
product: 
$$
 {\tilde {H}} = BA\;.
$$
Then $ {\tilde {H}}$ and $H$ have the same non zero eigenvalues. 
Indeed, if $\lambda$
is an eigenvalue $( \lambda \not= 0)$ of $ \tilde {H}$:
$$
\tilde {H} \mid {\tilde \Psi }_{\lambda }> = \lambda{\mid {\tilde  \Psi 
}_{\lambda }>} \;,
$$
then
$$
 \mid {\Psi }_{\lambda }> = A{\mid {\tilde \Psi}_{\lambda }>} \;,
$$
obeys:
$$ 
  H\mid {\Psi }_{\lambda }> = \lambda{\mid {\Psi}_{\lambda }>} \;.
$$ 
For the harmonic oscillator,
$$
H = (a^{\dagger } a + 1/2) \;,
$$
with  
$$  
[ a , a^{\dagger } ] = 1 \;,
$$
introduce:
$$
{\tilde  H } =  (a a^{\dagger }  + 1/2 ) =  (a^{\dagger } a + 3/2) \;.
$$
The operators $ H$ and $ \tilde  H $ are positive operators, and it 
follows that the vector 
$\mid {0}>$,
$$
a\mid {0}> = 0
$$
is an eigenvector of  $ H $, with $\lambda = 1/2$, and of 
$\tilde  H $, with $\lambda = 3/2$, and therefore
$$
a^{\dagger }\mid {0}> \;,  
$$
is  an eigenvector of  $ H $,  corresponding to  
 $\lambda = 3/2$. The procedure can be continued:
$$
\tilde {\tilde  H} = H^{(2)} = \tilde H + 1 = H + 2 \;.
$$
At the $n$-th step
$$
H^{(n)} = H + n \;,
$$
with $\mid {0}>$ being an eigenvector of  $ H^{(n)} $corresponding to
$$
\lambda =  n + 1/2 \;.  
$$ 
Then $ H $, has the same eigenvalue and the corresponding eigenvector 
is:  
$$
{\mid {\Psi}_{\lambda }>} = {(a^{\dagger })^{n}}\mid {0}> \;.
$$

We emphasize that the factorization method involves three steps: First, 
write the Hamiltonian in a factorized way; second, use a trick 
to cast $\tilde H $ in a form similar to that of $ H $ (i.e. move the 
destruction operators to the right in the previous example). The third 
step is to find the necessary recurrence procedure that yields all the 
eigenvalues and eigenvectors.  This is also the pattern we are going 
to follow.

In Section 2 we reformulate the method of reference \cite{fera} for the scalar
wave function in the einbein formalism. We show that the recurrence 
relations are obtained as a consequence of the requirement that the 
reverse order product $ BA $ has the same covariance properties acting 
on scalar wave vectors as the original product $ AB $. On the sphere 
this happens because the spin connection associated with the $ U(1) $ gauge 
group in the tangent space can be chosen to be proportional to the 
gauge connection corresponding to the monopole at its center\cite{edwit}, the 
proportionality factor  being exactly the number entering into 
the Dirac quantization condition.

In Section 3  we formulate the 
maximal classical supersymmetric action associated with the motion of 
a charge on a sphere, in the field of a monopole located at its center.
The same action can be obtained 
by considering the supersymmetric one dimensional $SU(2)/U(1)$  nonlinear
model and coupling it with the electromagnetic field through 
a gauge potential equal with the connection, generated by the nonlinear
transformation law\cite {pashnev}. The equivalence of these two 
formulations stems from the aforementioned property: the spin connection 
and the gauge connection are 
proportional. Alternatively, the same result  follows by  
restricting the $ N = 1$ supersymmetric action corresponding to a charge 
in the field of a monopole through Lagrange super-multipliers \cite 
{fdej} \cite {plyush}. However, we prefer to obtain our model from the 
existing general superspace $\sigma $-model actions \cite {papaone} 
\cite {papatwo}, by fixing the field content, and choosing the 
appropriate background. In this way we can follow the natural way of 
solving, models of this sort.  We will need only two 
real anticommuting classical degrees of 
freedom for the formulation of the classical supersymmetric 
action. After quantization these real anticommuting  degrees of freedom
become the gamma matrices. 
This suggests that the two bosonic degrees of freedom combine with 
the anticommuting ones in two type $B$ superfields. Then we fix the 
appropriate background in the most general renormalizable  $N =1 $ type $B$ 
superspace $\sigma $-model action\cite {papaone} \cite {papatwo}. 
The quantization of the general superspace $\sigma$-models produces the 
component approach to supersymmetric quantum mechanics, the 
general formulation of which, was pursued in \cite {ritt}. To quantize
our model we use the procedure of \cite{andy1} \cite{andy2} in which 
the reparametrization covariance can be obtained with the help of the 
appropriately defined Noether supercharge.  
 
Section 4 is devoted to the diagonalization of this quantum system,
while in Section 5 we give some concluding remarks. 
 
\section{ Factorization Method}

The Hamiltonian for the motion of a charge $ e $, on a sphere, in the 
field of a monopole of strength $ g $, localized in the center of the 
sphere is:
\eqn\label{H_{N}}
H_{N} = -{1\over 4} g^{ab}\nabla_{a}^{(N)}\nabla_{b}^{(N)} \;,
\e 
where  $ g^{ab} $ is the inverse metric on a sphere:
\eqn\label{d s^{2}}
d s^{2} = g_{ab} dx^{a}dx^{b} = R^{2} {\sin }^{2} \theta d^{2}\phi + 
R^{2}d^{2}\theta \;, 
\e
and the covariant derivatives are:
\eqn\label{nabla}
\nabla_{b}^{(N)} = { \partial }_{b} -{ \dot {\imath} } A_{b}^{(N)}\;,
\e 
and:
\eqn\label{2nabla}
\nabla_{a}^{(N)}\nabla_{b}^{(N)} =  { \partial }_{a}\nabla 
_{b}^{(N)}
 -  \dot {\imath} A_{a}^{(N)}\nabla_{b}^{(N)} - \Gamma^{c}_{ab}
\nabla_{c}^{(N)}\;, 
\e 
where $  { \partial }_{a} $ are the derivatives with respect to the 
coordinates in a patch on the sphere. The gauge connection  $ A_{a}^{N} $
is:
\eqn\label{Gaugeconn}
A_{\phi } = {(eg)\over {R\sin \theta }} (\pm 1 - \cos \theta ) \;,
\e
where $ \theta < \pi $ for the upper sign and $ \theta > 0 $  for the 
lower sign.  $ \Gamma^{c}_{ab} $ is the standard Cristoffel 
connection for a sphere.

The desired factorization of (\ref{H_{N}}) will be obtained by 
introducing the stereographic projection:
\eqn\label{Stereo}
{cos {\theta}}   = {1 - {2\over{1 + {{x^{2} + y^{2}}\over {4R^{2}}}}}}\;,
\e

\eqn\label{azi}
{\tan {\phi}} = {y\over x }\;.
\e
Defining the complex coordinate:
\eqn\label{z}
z = x +  \dot {\imath} y \;,
\e
the  only nonvanishing components of $ g^{ab} $, are  $ g^{z \bar z} = 
g^{{\bar z} z } $ with:
\eqn\label{g}
g^{z \bar z} = {2} {h^{-2}} = 2({1 + {{ z\bar z}\over {4R^{2}}}}) \;.
\e
Now, in the new coordinates the Hamiltonian is:
\eqn\label{H}
H_{N} = -{1\over 2}g^{z \bar z} \nabla_{\bar z }^{(N)} 
\nabla_{z}^{(N)} + 
{N\over {4R^{2}}}\;,
\e
where
\eqn\label{nablac}
\nabla_{z}^{(N)} = {\partial}_{z} - \dot {\imath} A^{(N)}_{z}\;,
\e
\eqn\label{nablac1}
\nabla_{\bar z}^{(N)} = {\partial}_{\bar z} -  \dot {\imath} A_{\bar 
z}^{(N)}\;,
\e
and
\eqn\label{cconn}
A_{z}^{(N)} = {1\over 2}(A_{1}^{(N)} -  \dot {\imath} A_{2}^{(N)})  = 
-\dot {\imath} N{\partial}_{z} \ln h \;\;\;,\;\;\; 
A_{\bar z}^{(N)} = {A_{z}^{(N)}}^{\ast }\;.
\e
The Cristoffel connection does not appear in (\ref{H}) anymore, because
its only nonvanishing components are  $ \Gamma^{z}_{zz}$and $
 \Gamma^{\bar z}_{\bar z \bar z} $ .
In (\ref{H}) we have used the relation:
\eqn\label{com} 
{g^{z \bar z}}[\nabla_{z}^{(N)} , \nabla_{\bar z}^{(N)}]
 = - {{N}\over{R^{2}}}\;,
\e
in order to exhibit the ``destruction operators'' ${{\nabla 
}_{z}}^{(N)}$ to the right and the 
``creation operators'' $\nabla_{\bar z}^{(N)}$ to the left. 
The eigenvalues of  $H_{N}$ are defined by the zero modes of the 
operators  $\nabla_{z}^{(K)}  $ . These zero modes are related 
by the index theorem to topological properties of the manifold, 
see reference \cite{fera}. Thus we may regard the present method of 
integrating $ H_{N} $  as a topological one.

In our case the wave vector is a scalar under the 
reparametrizations  of the manifold. In a more general setting we 
might consider different assignments of spinorial (tensorial) 
properties of the wave vector. This is the case when one deals with the 
motion a charged spin one half particle on a sphere, when the wave 
function is a spinor. Then, for a manifestly covariant approach, the 
appropriate language is the vielbein formalism of general relativity. 
This is not actually necessary for the case of the scalar wave 
functions. However, as we will show, even in this case, it allows one 
to avoid the use of the correct, but rather 
artificial, similarity transformations  \cite{fera}  in establishing the 
recurrence relations necessary to apply the factorization method.

Let us now introduce the ein-beins for our complex manifold:
\eqn\label{ ein}  
g_{z \bar z} = e_{z}^{+}e_{\bar z}^{-} {\eta}_{+-}\;,
\e
where  $\eta_{+-} $ is the metric in tangent space in an appropriate 
basis:
\eqn\label{eta}
 {{\eta}_{+}}_{-} = {\eta}_{-+} = {1\over 2}  \;\;\;,{{\eta}^{+}}^{-} = 
{\eta}^{-+} =  2 \;.
\e
We have:
\eqn\label{ defe}
e_{z}^{+} = e_{\bar z}^{-} = h  \;\;\;,\;\;\; e_{z}^{-} = 
e_{\bar z}^{+} = 0\;.
\e

Using the constant covariance of the ein-bein:
\eqn\label{ cove}
\nabla_{a}e_{b}^{\alpha} = \partial_{a}e_{b}^{\alpha } + {{{\omega 
}_{a}}^{\alpha}}_{\beta }e_{b}^{\beta } - \Gamma^{c}_{ab}
e_{c}^{\alpha } = 0\;,
\e
 (\ref {H}) becomes:
\eqn\label{Hprime}   
H_{N} = -{\mathcal D}_{-}^{(N)} {\mathcal D}_{+}^{(N)} + 
{N\over {4R^{2}}}\;,
\e 
where:
\eqn\label{scriptp}
{\mathcal  D}_{-}^{(N)} = {e_{-}^{\bar z}}(\nabla_{\bar z}^{(N)}
 + {{{\omega }_{z}}_{+}}^{+} )\;,
\e 
and 
\eqn\label{scriptm}
{\mathcal  D}_{+}^{(N)} = {e_{+}^{ z}} \nabla_{z}^{(N)}\;.
\e
Above, one has:
\eqn\label{ defei}
e_{+}^{z} = e_{-}^{\bar z} = {h}^{-1}\;,
\e
and the nonvanishing components of the spin connection are:
\eqn\label{spinconnection1}
{{{\omega}_{z}}^{+}}_{+} = -{{{\omega}_{z}}_{+}}^{+} = 
{\partial}_{z}\ln h\;,
\e  
\eqn\label{spinconnection2}
{{{\omega}_{\bar z}}^{+}}_{+} = -{{{\omega}_{\bar z}}_{+}}^{+} = -
{\partial}_{\bar z}\ln h\;.
\e   
Consider now the tilde of (\ref{Hprime}) : 
\eqn\label{Htil}
{\tilde {H}}_{N} = -[{\mathcal  D}_{+}^{(N)}]_{nc}[{\mathcal  
D}_{-}^{(N)}]_{nc} + 
{N\over {4R^{2}}}\;,
\e
While  (\ref {Hprime}) is manifestly generally covariant ( assuming that 
the wave vector is a world scalar), (\ref {Htil}) is not manifestly so.
By reversing 
the order of covariant derivatives, the spin connection terms in the 
covariant derivatives do not match the tensor properties of the 
terms ahead of them. This is why in considering the reverse order product 
above, we appended the $nc$ index to the covariant derivatives, even if their 
definition (\ref {scriptp}), (\ref {scriptm}) did not change. 

In order to obtain a 
recurrence relation one would expect (as explained in the 
Introduction) the operator $ \tilde {H}_{N} $ to be a scalar and act 
on a scalar wave function. However, then  $ \tilde {H}_{N} $  is not manifestly 
covariant. The manifest covariance of  $ \tilde {H}_{N} $ is restored by noting 
that the gauge connection and the spin connection can be chosen to 
be proportional \cite{edwit}:
\eqn\label{Prop}
A_{a}^{(N)} = N{\omega }_{a}\;,
\e
where $ N = eg $ appears in the Dirac quantization condition. 
Therefore we have the following identity:
\begin{eqnarray} 
[{\mathcal  D}_{+}^{(N)}]_{nc}[{\mathcal  
D}_{-}^{(N)}]_{nc} = e_{+}^{z}\nabla_{z}^{(N)}e_{-}^{\bar z}
(\nabla_{\bar z}^{(N)}+{{{\omega }_{z}}_{+}}^{+}) = \nonumber \\
 \; \nonumber \\ 
e_{+}^{z}(\nabla_{z}^{(N+1)} + {{{\omega }_{z}}_{-}}^{-})e_{-}^{\bar z}
\nabla_{\bar z}^{(N+1)} = {\mathcal  D}_{+}^{(N + 1)}
{\mathcal  D}_{-}^{(N + 1)} \;.
\end{eqnarray}
Here the expression of  $ {\mathcal  D}_{+}^{(N + 1)}$
,  $ {{\mathcal  D}_{-}^{(N + 1)}}$, can be read in the above 
formula and the the expression to the right, above, is fully covariant. 
Substituting this in  $ \tilde {H}_{N} $ and moving the destruction 
operators to the right with the help of  (\ref {com}) we have: 
\eqn\label{Prop1}
{\tilde {H}}_{N} = {H_{N}}^{(1)} =  H_{N+1} + {(2N+1)\over {4R^{2}}}\;.
\e
Therefore the next eigenvalue of $ {H}_{N} $ is  $ {3N+2}\over {4R^{2}} $.
The corresponding eigenvector can be obtained from that of 
$ \tilde {H}_{N} $, given by the condition
\eqn\label{eigen}
\nabla_{z}^{(N+1)}{\tilde \Psi }_{1} = 0\;.
\e
One has:
\eqn\label{eigenv1}
{\Psi }_{1} = {\mathcal D }_{-}^{(N)}{\tilde \Psi }_{1}\;,
\e
with $ {\mathcal  D}_{-}^{(N )} $ defined by (\ref {scriptp}).
Even if we deal with two covariant problems, that of $ {H}_{N} $ and 
that of $ \tilde {H}_{N} $, the connection between the two sets of eigenvectors 
is not generally covariant.

The procedure described above can be continued, and at the $l-th$ step 
we get 
\eqn\label{Hl}
 H_{N}^{(l)} =  H_{N+l} + {[(2N - 1) +l(l+1) ]\over {4R^{2}}}\;,
\e
with the eigenvalue
\eqn\label{leigenvalue}
{\lambda }_{l} = {1\over {4R^{2}}}[ (2l + 1) N + l(l+1) ]\;,
\e 
anthe the corresponding eigenvector of $ {H}_{N} $ 
\eqn\label{eigenv2}
{\Psi }_{l} = {\mathcal D }_{-}^{(N)} \ldots {{\mathcal D 
}_{-}^{(N+l-1)}}{\tilde \Psi }_{l}\;,
\e
where $ {{\mathcal  D}_{-}^{(P)}}  $, was defined in (\ref {scriptp}).
${\tilde \Psi }_{l}$, is the solution of
\eqn\label{zeromodel}
\nabla_{z}^{(N+l)}{\tilde \Psi }_{l} = 0\;.
\e
The multiplicity of the state $ {\Psi }_{l}$ is obtained from the 
condition of finite norm of the states:
\eqn\label{norm}
{\int \nolimits } {{dz d{\bar z}}\over 2} h^{2} {\vert {{\tilde \Psi}}_{l}
\vert}^{2}< \infty \;,
\e
Indeed, the solution of (\ref {zeromodel}) is:
\eqn\label{sol}
{\tilde \Psi }_{l} = {h}^{N+l}f(\bar z)\;,
\e
Where $ f(\bar z) $ is an arbitrary polynomial of degree  $ \leq 
2(N+l) $, making the degeneracy of the corresponding state $ 2(N + 
l) + 1 $ . From the asymptotic behaviour of (\ref {eigenv2}) in the 
radial variable, one sees that there are potential problems with 
normalizability of such states. However, as  we checked on examples,   
due to cancelations of unwanted terms the vector $ {\Psi }_{l} $ is 
normalizable. We point out that this result is valid for integer 
$ 2N \geq 0 $. Otherwise $ (N < 0) $ the creation and annihilation operators
must be interchanged.  

Hence imposing the manifest general covariance of $ \tilde {H}_{N} $ 
led us to  rederive the  recurrence relations  \cite {fera} necessary in 
order to completely integrate the Hamiltonian $ {H}_{N} $. In the next 
section using the canonical quantization we will obtain the 
Hamiltonian for the supersymmetric particle which will be 
subsequently diagonalized by factorization method.  


\section{$ N= 2$ Supersymmetric Quantum Mechanics on a Sphere}

We will approach the supersymmetrization of a given bosonic action in 
the following way. Given the target manifold (whose local coordinates 
are the bosonic fields which appear in the formulation of the 1-dimensional 
$\sigma$-model of  the system), we look at the dimensionality of the Clifford 
algebra supported by the tangent space, for the sphere this is two. 
Therefore in the present case  the $ \Gamma $-matrices will be hermitian 
matrices and therefore can be obtained from the quantization of the two 
real anticommuting degrees of freedom. Thus the minimal content of 
the fields realizing the representation of the supersymmetry algebra will
be two real bosonic (the coordinate on the target manifold ) and two 
real anticommuting degrees of freedom. We can fit this degrees of 
freedom in two type-$ B $ superfields. Because the tangent space is 
two dimensional and the supersymmetry charges must be constructed with 
the help of the  $ \Gamma $-matrices, one expects the maximal 
supersymmetry allowable for the system to be $ N =2$. Thus with the 
help of two type-$ B $ superfields we must formulate an  $ N =2$ 
supersymmetric action. This is  automatic since our target space 
manifold admits a complex structure. The above argument is somehow 
circular because one formulates the problem on the basis of the 
outcome of the quantization procedure.

Therefore with the help of the metric  $ {g_{a}}_{b}  $ , the gauge
connection ${A_{a}}^{(N)} $ and the type-$B$ superfield   $ X^{a}(x , 
\theta) $ , we construct the following $ N = 1$, 1-dimensional  sigma-model
\cite {papaone} \cite{papatwo} :
\eqn\label{Action} 
S = - \dot {\imath} {\int \nolimits } dt d\theta {\bigl \lbrace } 
{{{g_{z}}_{\bar z}\over 2} {\bigl ( } DX^{z} {\dot X}^{\bar z} +
DX^{\bar z} {\dot X}^{z}{\bigr ) } + A_{z} DX^{z} + A_{\bar z}DX^{\bar 
z}{\bigr \rbrace } } \;. 
\e 
Here
\eqn\label{covders}
D = {\partial \over {\partial \theta } } +  \dot {\imath} \theta {d \over dt}\;,
\e 
is the covariant supersymmetric derivative. The superfields $ X^{z} $ 
and  $ X^{\bar z} $, are connected through 
 $ X^{\bar z} = (X^{z})^{\dagger } $ and have the components:
\eqn\label{components}
z = {X^{z}\vert}_{\theta = 0} \;\;\;\;\;, \;\;\;\;\; {\lambda }^{z} = {( DX^{z}
)\vert}_{\theta = 0}\;,
\e 
with:
\eqn\label{conjug1}
{{\lambda }^{z}}^{\dagger } = - {\lambda }^{\bar z}\;,
\e 
and
\eqn\label{conjug2}
{\int \nolimits } d\theta { \lbrace } \ldots { \rbrace }  = D 
{ \lbrace } \ldots { \rbrace }{\vert_{\theta = 0}}\;.
\e 
The action   (\ref {Action}) is invariant under the supersymmetry 
transformations
\eqn\label{susy}
{\delta}_{\epsilon } X^{\bar z} = \epsilon Q  X^{\bar z}\;,
\e 
\eqn\label{susybar} 
{\delta}_{\epsilon } X^{z} = \epsilon Q  X^{z}\;,
\e 
where the supersymmetry shift operator $ Q $ is
\eqn\label{Q}
Q = {\partial \over {\partial \theta } } -  \dot {\imath} \theta {d \over dt}\;,
\e  
with
\eqn\label{ALG}
Q^{2} = - \dot {\imath} {d\over dt}\;\;\;\; , \;\;\;\;  \lbrace Q, D \rbrace = 0\;,
\e  
As mentioned before, because of the fact that the target space 
manifold is complex we will automatically have a second supersymmetry 
\eqn\label{susy2}
{\delta}_{\eta  } X^{\bar z} = - \dot {\imath}  \eta   D  X^{\bar z}\;,
\e 
\eqn\label{susybar2} 
{\delta}_{\eta   } X^{z} =  \dot {\imath} \eta  D  X^{z}\;.
\e
The two supersymmetry transformations above can be combined in one 
complex supersymmetry which in component fields $ z, \bar z, 
{\lambda }^{z} , {\lambda }^{\bar z} $ takes the form:
\eqn\label{susycomplex1}
\delta z = \delta \mu {\lambda }^{z}\;\;\;\; ,\;\;\;\;\delta{\lambda 
}^{z} =  \dot {\imath}  \delta \bar \mu \dot z \;,
\e
and 
\eqn\label{susycomplex2}
\delta {\bar z} = \delta {\bar \mu} {\lambda }^{\bar z}\;\;\;\; 
,\;\;\;\;\delta{\lambda 
}^{\bar z} =  \dot {\imath}  \delta \mu \dot {\bar z} \;.
\e
Where 
\eqn\label{bar3}
\delta \mu ^{\dagger } = \delta {\bar \mu} \;.
\e

The action (\ref {Action})  can also be written in terms of component 
fields, and  the Lagrangian  is:
\eqn\label{lagrangeian}
{\mathcal  L} = g_{z\bar z} {\dot z} {\dot {\bar z}} +  \dot {\imath} { g_{z\bar z} 
\over 2} ( {\lambda }^{z} {\sc D} {\lambda }^{\bar z} + {\lambda 
}^{\bar z} {\sc D} {\lambda }^{z} ) -  \dot {\imath} F^{(N)}_{z\bar 
z}{\lambda}^{z}{\lambda}^{\bar z} 
+ A_{z}^{(N)}{\dot z}+ A_{\bar z}^{(N)}{\dot{\bar  z}} \;,
\e
where
\eqn\label{streghzc}
{{F_{z}}_{\bar z}} = {\partial}_{z}A_{\bar z} - {\partial}_{\bar z}A_{z}\;,
\e
and 
\eqn\label{pullb}
D{\lambda }^{z} = {\dot \lambda }^{z} + {\dot z}\Gamma^{z}_{zz}
{\lambda }^{z}\;\;\;\;\;\; , \;\;\;\;\;D{\lambda }^{\bar z} = {\dot \lambda }
^{\bar z} + {\dot {\bar z}}\Gamma^{\bar z}_{{\bar z}{\bar z}}
{\lambda }^{\bar z}\;,
\e
where  $\Gamma^{z}_{zz}$  and $\Gamma^{\bar z}_{{\bar z}{\bar z}}$
are the only nonvanishing components of the Cristoffel connection for 
the sphere.  (\ref {lagrangeian}) is invariant under (\ref {susycomplex1}) 
and (\ref {susycomplex2}) and by the Noether procedure one can deduce 
the corresponding supercharges:
\eqn\label{superq}
\bar Q = -{\lambda^{z}}{g_{z \bar z}}{\dot {\bar z}}\;,
\e
\eqn\label{superq1}
Q = {\lambda^{\bar z}}{g_{{\bar z}z}}{\dot {z}} \;.
\e
>From (\ref {lagrangeian}) one infers that one can go to the tangent 
space indices by making the redefinition :
\eqn\label{redefla}
{\lambda}^{z} =  \dot {\imath} e_{+}^{z}{\lambda}^{+} \;,
\e 
\eqn\label{redeflaba}
{\lambda}^{\bar z} =  \dot {\imath} e_{-}^{\bar z}{\lambda}^{-} \;.
\e 
Then the fermions in (\ref {lagrangeian}) acquire a standard form:
\eqn\label{redeflag}
{\mathcal L} = {g_{z}}_{\bar z} {\dot z}{\dot {\bar z}} - { \dot {\imath} \over 
4}({\lambda}^{+}{\dot {\lambda}}^{-} + {\lambda}^{-}{\dot {\lambda}}^{+})
+ \ldots \;,
\e
and one concludes that $ {\lambda}^{\pm } $ quantizes as the 
corresponding $ {1\over {\sqrt 2}} {\Gamma }^{\pm } $ matrices:
\eqn\label{gammaplmin}
{\Gamma}^{(+)} = {{\Gamma}^{(-)}}^{\dagger } = 
\left( \begin{array}{cc}
   0   &   2\\
   0   &   0
   \end {array} 
   \right ) \;,
\e 
The canonical conjugate momentum is given by:
\eqn\label{canmom} 
{\sc {P}}_{z} = g_{z\bar z}{\dot {\bar z}} + { \dot {\imath}\over 
2}({{\omega}_{z}})_{\alpha \beta}{\lambda}^{\alpha}{\lambda}^{\beta} +
A_{z}^{(N)}\;.
\e 
With the notation:
\eqn\label{covsupc} 
{\sc {\Pi }}_{z} = g_{z\bar z}{\dot {\bar z}} = {\sc {P}}_{z} - { \dot {\imath}\over 
2}({{\omega}_{z}})_{\alpha \beta}{\lambda}^{\alpha}{\lambda}^{\beta}  -
A_{z}^{(N)}\;,
\e 
the corresponding Noether charge can be written :
\eqn\label{qsuperch} 
 Q ={\lambda }^{z}{\sc {\Pi }}_{z} =  \dot {\imath} e_{+}^{z}{\lambda }^{+}
 ({\sc {P}}_{z} - { \dot {\imath}\over 
2}({{\omega}_{z}})_{\alpha \beta}{\lambda}^{\alpha}{\lambda}^{\beta}  -
A_{z}^{(N)})\;.
\e 
 Let us now quantize the supercharge $Q$ \cite {andy1} \cite {andy2}.
 As mentioned before ${\lambda}^{\pm}$ quantizes as the 
 corresponding  $ {1\over {\sqrt 2}} {\Gamma }^{\pm } $ , further
 $ {\sc {P}}_{z} $ goes into ${1\over { \dot {\imath}}}{\partial \over 
 {\partial z}}$, therefore in order to maintain the general 
 covariance under quantization we can take the minimal $Q$:
\eqn\label{qsuperchmin}
Q = {{{\Gamma }^{+}}\over {\sqrt 2}}e_{+}^{z}({1\over { \dot {\imath}}})
\nabla_{z}^{(N)}\;,
\e 
where $ {{\nabla}_{z}}^{(N)} $, is the covariant derivative, on the 
spinor wave function:
\eqn\label{qcovderq}
\nabla_{z}^{(N)} = {\partial}_{z} - \dot {\imath}
 A_{z}^{(N)} -
{1\over 4}{{{\omega}_{z}}^{\alpha \beta}}{\Gamma }_{\alpha \beta }\;,
\e 
and the matrices ${\Gamma }_{\alpha \beta} $ are
\eqn\label{gammaabs}
{\Gamma }_{\alpha \beta} = {1\over 2} [{\Gamma}_{\alpha } , 
{\Gamma}_{\beta } ]\;,
\e 
and
\eqn\label{anticgammaabs}
{\lbrace {\Gamma}_{\alpha } , {\Gamma}_{\beta } \rbrace} = 2 {\eta 
}_{\alpha \beta }\;,
\e  
Once we have the quantum expression for $Q$ we can define
\eqn\label{qbarq}
{\bar Q} = {Q}^{\dagger}\;,
\e 
with respect to the scalar product corresponding to (\ref 
{norm}), where of course now the scalar wave vector is replaced by 
the two component wave function. The adjoint of $Q$ is:
\eqn\label{qbarsuperchmin}
{ Q^{\dagger }} = -{{{\Gamma }^{-}}\over {\sqrt 2}}e_{-}^{\bar z}
\nabla_{\bar z}^{(N)}\;,
\e
with               
\eqn\label{qbarcovderq}
\nabla_{\bar z}^{(N)} = {\partial}_{\bar z} -  \dot {\imath} A_{\bar 
z}^{(N)} -
{1\over 4}{{{\omega}_{\bar z}}^{\alpha \beta}}{\Gamma }_{\alpha \beta }\;,
\e
The $Q$ and $\bar Q $ so defined obey automatically
\eqn\label{idempotence}
Q^{2} = {\bar Q}^{2} = 0\;,
\e
because they contain the matrices $ {\Gamma}^{\pm } $ in their 
definition. 
Defining the quantum Hamiltonian by 
\eqn\label{qhamilt}
H = {1\over 2} ( Q\bar Q + {\bar Q} Q )\;,
\e
$H$ commutes automatically with the supercharges $Q$ and $\bar Q$.
Thus the quantum ordering in $H$ is completely fixed by 
supersymmetry and reparametrization covariance. The expression for $H$
is also covariant under the 
reparametrizations of the manifold and the system has $ N = 2 $ 
supersymmetry.

Finally it might be worth mentioning that by this procedure $H$ is 
defined from the supersymmetry algebra  (\ref {qhamilt}) without any 
recourse to the standard ways of defining it.

\section{ Supersymmetric Monopole Harmonics}

Diagonalization of the Hamiltonian (\ref {qhamilt}) follows now in a
rather simple way. Using the constant covariance of the $ 
\Gamma$-matrices with respect to vector and spinor indices (\ref {qhamilt})
can be cast in the following form
\eqn\label{susyhamN}
H = - {1\over 4}e_{+}^{z} e_{-}^{\bar z}[ {{\Gamma }^{+}}{{\Gamma }^{-}} 
 \nabla_{z}^{(N)}\nabla _{\bar z}^{(N)} +
{{\Gamma }^{-}}{{\Gamma }^{+}}\nabla_{\bar z}^{(N)}\nabla 
_{z}^{(N)}]\;,
\e
the products $ {{\Gamma }^{+}}{{\Gamma }^{-}} $ and  $ {{\Gamma }^{-}}
{{\Gamma } ^{+}} $ are scalars under local rotations in the tangent 
plane. Moreover they are projectors:
\eqn\label{gammaproject}
{\Gamma}^{(+)}{{\Gamma}^{(-)}} = 
\left( \begin{array}{cc}
   4   &   0\\
   0   &   0
   \end {array} 
   \right ) \;\;\;\;\;\;, \;\;\;\;\;
   {\Gamma}^{(-)}{{\Gamma}^{(+)}} = 
\left( \begin{array}{cc}
   0  &    0\\
   0   &   4
   \end {array}
   \right ) \;,
\e
Therefore $H$ is a sum of factorized terms. We recognize that the 
operators multiplying the products   $ {{\Gamma }^{+}}{{\Gamma }^{-}} $ and 
$ {{\Gamma }^{-}}{{\Gamma }^{+}} $   are connected with the 
Laplacian appearing in (\ref {H_{N}}) with a modified spin 
connection, due to the nonzero spin of the wave function. In fact 
we have:
\eqn\label{hamiltsp}
H =[ H_{(N+{1\over 2})} + {{( N + {1\over 2} )}\over {4R^{2}}} ]
{{\Gamma}^{(+)}{{\Gamma}^{(-)}}\over 4} +
[ H_{(N - {1\over 2})} - {{( N - {1\over 2} )}\over {4R^{2}}} ]
{{\Gamma}^{(-)}{{\Gamma}^{(+)}}\over 4}\;,
\e
This is due to the fact that the spinor components transform with an 
effective charge $\pm {1\over 2}$ under local rotations. As remarked 
before one can absorb this ``Lorentz charge'' in the gauge connection leading 
to a modification of the effective charge of the corresponding components.
Using  (\ref{Prop1}), it is easy to show that:  
\eqn\label{tildehamilthsp} 
{{\tilde H}_{(N - {1\over 2})}} - {{( N - {1\over 2} )}\over {4R^{2}}}  = 
 H_{(N+{1\over 2})} + {{( N + {1\over 2} )}\over {4R^{2}}}\;,
\e   
However,  by the same procedure  $ {{\tilde H}_{N + {1\over 2}}} $  gets 
connected with $  {H_{N + {3\over 2}}} $. The connection between
 ${ {\tilde H}_{N + {1\over 2}}} $ and  $  {H_{N -{1\over 2}}} $  is the
 result of a different factorization of $ H_{N} $ in Section 2. From  
 (\ref{H_{N}}) one has:
\eqn\label{Hprimesec}   
H_{N} = -{\mathcal D}_{+}^{(N)} {\mathcal  D}_{-}^{(N)} - 
{N\over {4R^{2}}}\;,
\e 
where
\eqn\label{scriptpsec}
{\mathcal  D}_{+}^{(N)} = e_{+}^{z}( \nabla_{z}^{(N)}
 + {{{\omega }_{z}}_{-}}^{-} )\;,
\e 
and 
\eqn\label{scriptmsec}
{\mathcal  D}_{-}^{(N)} = {{e_{-}^{\bar  z}} \nabla_{\bar z}^{(N)}}\;,
\e  
therefore with respect to this factorization we have:
\begin{eqnarray}  
{\tilde {H}}_{N} = -[{\mathcal  D}_{-}^{(N)}]_{nc}[{\mathcal  
D}_{+}^{(N)}]_{nc} - {N\over {4R^{2}}} = \nonumber  \\
\; \nonumber  \\
-{\mathcal D}_{-}^{(N-1)} {\mathcal  
D}_{+}^{(N-1)} - {N\over {4R^{2}}} = H_{N-1} - {{2N-1}\over {4R^{2}}}\;,
\end{eqnarray}
and using the identities just derived one has:
\eqn\label{tildehamilthspsec} 
{{\tilde  H}_{(N + {1\over 2})}} + {{( N + {1\over 2} )}\over {4R^{2}}}  = 
 H_{(N-{1\over 2})} - {{( N - {1\over 2} )}\over {4R^{2}}}\;,
\e  
We prefer to start with the eigenfunctions of $ {H_{N + {1\over 
2}}} $ , because this operator appears in a manifestly positive 
definite combination in the Hamiltonian.

Therefore given the nonzero eigenvalue eigenvector $\Psi $ of
 $  H_{N+{1\over 2}} + ( N + {1\over 2}) $ we obtain the corresponding
eigenvector of $ H_{N - {1\over 2}} - ( N - {1\over 2}) $ by 
taking: 
\eqn\label{eigevesp}
{\mathcal D}_{-}^{(N - {1\over 2})} {\Psi} \;,
\e
with $ {\mathcal  D}_{-}^{(N - {1 \over 2})}  $, from (\ref {scriptp})
>From (\ref{hamiltsp}) the eigenvalues of  $ H $ are:
\eqn\label{eigenvalspin}
E_{l} = {1\over{4R^{2}}}(l+1)[l+2N+1]\;,
\e
with the eigenvector being given by
\eqn\label{eigevesptot}
{\Psi }_{l} =
\left( \begin{array}{c}
{{\mathcal    D }_{-}^{(N+{1 \over 2})}} \ldots 
{{\mathcal  D }_{-}^{(N + l - {1 \over 2})}}
{\tilde \Psi }_{l}\\
 {\mathcal D}_{-}^{(N- {1 \over 2})}{{\mathcal   D }_{-}^{(N+{1 \over 2})}}
  \ldots {{\mathcal   D }_{-}^{(N+l - {1 \over 2})}}
{\tilde \Psi }_{l}                   
   \end {array} 
   \right ) \;.
\e 
Therefore we have found the eigenvalues and eigenvectors of the  $ N= 
2$  supersymmetric spinning particle moving on a sphere in the field 
of a monopole. In spherical coordinates they appear in \cite {lucv}.

\section{Conclusions}

   Restating the main result, we have supersymmetrized and solved the 
motion of a charge on a sphere in the field of a monopole at its 
center.

The factorization method appears to be the natural way to solve this 
problem,. This is because  in order to formulate a supersymmetric 
problem one is basically compelled to use the complex structure of 
the target space manifold. The quantization scheme for the problem is 
manifestly taking into account the reparametrizations of the manifold 
therefore it is covariant with respect to this reparametrizations. One 
should remember that one is dealing basically with the algebra of the 
angular momentum and it is quite interesting that following the 
manifest symmetries of the problem one is led to an alternative 
integration method. In this context, even if algebraic, this method 
appears somewhat strange, albeit natural.

One should also stress that the degeneracies of the levels are all 
finite and it is well known that we deal with a regularization of the 
Landau electrons.  
Taking the the limit $ R^{2}, N \rightarrow \infty $ (with $ N\over R^{2} $ 
fixed) one obtains the infinetely degenerate states of a planar 
electron in a constant magnetic field.  

In a rather different context, the eigenfunctions obtained in this 
paper may help to define an alternative harmonic superspace \cite {jenia}.
 
\vspace{0.3in}
{\bf Acknowledgments}
We wish to thank Willy Fischler for the question which led to this 
problem and for subsequent discussions. We have also benefited 
from discussions with S. Barnes, E.A. Ivanov, A.I. 
Pashnev, V. Rittenberg and A.B Zamolodchikov. G.A.M. gratefully
ackowledges financial support from a grant by Army Research Laboratory, 
BAE Systems.  L.M. wishes to thank E. Paschos for hospitality at the 
University of Dortmund, while this work was in progress and is  also 
indebted to A.T. Filippov for making possible his participation to 
the Symposium of Quantum Gravity and Superstrings, 18-28 June 2001, 
Dubna, where part of this work was presented. L.M. was supported 
in part by the National Science Foundation under grant PHY-9870101. 
\pagebreak

\end{document}